# Theoretical treatment of Coriolis effect using hyper-spherical coordinates, with application to the ro-vibrational spectrum of ozone

Igor Gayday,[*] Alexander Teplukhin,[†] Brian K. Kendrick[†] and Dmitri Babikov[*,1]

[*]*Department of Chemistry, Wehr Chemistry Building, Marquette University, Milwaukee, Wisconsin 53201-1881, USA*
[†]*Theoretical Division (T-1, MS B221), Los Alamos National Laboratory, Los Alamos, New Mexico 87545, USA*

## Abstract

Several alternative methods for description of interaction between rotation and vibration are compared and contrasted using hyper-spherical coordinates for a triatomic molecule. These methods differ by the choice of *z*-axis and by the assumption of a prolate or oblate rotor shape of the molecule. For each case, block-structure of the rotational-vibrational Hamiltonian matrix is derived and analyzed, and the advantages and disadvantages of each method are made explicit. This theory is then employed to compute ro-vibrational spectra of singly substituted ozone; roughly, 600 vibrational states of $^{16}O^{18}O^{16}O$ and $^{16}O^{16}O^{18}O$ isomers combined, with rotational excitations up to $J = 5$ and both inversion parities (21600 coupled ro-vibrational states total). Splittings between the states of different parities, so called *K*-doublings, are calculated and analyzed. The roles of the asymmetric-top rotor term and the Coriolis coupling term are determined individually, and it is found that they both affect these splittings, but in the opposite directions. Thus, the two effects partially cancel out, and the residual splittings are relatively small. Energies of the ro-vibrational states reported in this work for $^{16}O^{18}O^{16}O$ and $^{16}O^{16}O^{18}O$ are in excellent agreement with literature (available for low vibrational excitation). New data obtained here for the highly excited vibrational states enable the first systematic study of the Coriolis effect in symmetric and asymmetric isotopomers of ozone.

---

[1] Author to whom all correspondence should be addressed; electronic mail: dmitri.babikov@mu.edu

## I. INTRODUCTION

Accurate quantum mechanical treatment of coupled rotational-vibrational motion can be a challenging task, even for the smallest molecules such as triatomic, if the range of rotational and vibrational excitations is significant (*e.g.*, up to the dissociation threshold), the atoms are heavy (non-hydrogen) and the nature of problem requires accounting for all terms in the ro-vibrational Hamiltonian operator. With rotation-vibration interaction terms included, the size of the Hamiltonian matrix is usually huge, and the numerical cost of its diagonalization is very significant, often unpractical. In the literature such nearly exact calculations of the rotational-vibrational spectra have been reported for $H_3^+$,[1] $HeHF$,[2] $LiNC$,[3] $HeN_2^+$,[4] $H_2O$,[5] $H_2S$,[6] $SO_2$,[7,8] $HO_2$,[9] $Ar_3$,[10] and very recently for $O_3$.[11] For an accurate ro-vibrational spectrum like that, even after it has already been computed, the process of assigning the vibration mode quantum numbers ($v_1$, $v_2$, $v_3$) and the asymmetric-top rotor quantum numbers ($J_{KaKc}$) to the individual rotational-vibrational states, is also challenging.[12]

For these reasons, a symmetric-top rotor approximation (also known as the *K*-conserving assumption) remains a popular practical tool for the prediction of ro-vibrational state energies.[13,14,15,16] In this simplified method, the terms in the Hamiltonian operator, responsible for the coupling of rotational and vibrational degrees of freedom, are neglected (assumed to be small), which permits to split the overall Hamiltonian matrix into a number of independent smaller blocks that can be labeled by quantum numbers of the symmetric-top rotor ($J_K$). Within each block, accurate calculations of the vibrational states can be carried out, and then the overall spectrum of molecule is obtained by collating these individual pieces back together. The major drawback of this simplified approach is that the resultant spectrum lacks the so-called *K*-doubling.[7,17] Namely, for all values of *K* in the range $1 \leq K \leq J$, the ro-vibrational states computed in this simplified way are doubly-degenerate, while in nature they are known to exhibit non-zero splittings, the *K*-doubling.[7,17,18] Importantly, such splittings represent a unique spectroscopic feature of the molecule,[19] and may also play a role in natural phenomena, such as absorption of solar light by atmospheric species.[20,21]

Our interest in this methodological challenge was stimulated by the famous mystery of non-mass dependent fractionation of oxygen isotopes, associated with the recombination reaction that forms ozone,[22,23] which still remains unresolved.[24,25,26] One hypothesis, proposed by the group of Rudolf Marcus in a series of recent papers,[27,28] is that the Coriolis effect, responsible for the rotation-vibration interaction, occurs more efficiently in the isotopically substituted asymmetric ozone molecules, such as $^{16}O^{16}O^{18}O$. The group of Marcus carried out classical trajectory simulations to gain some insight into the mechanism of this phenomenon but did not find enough evidence for its justification.[28] Interestingly, they concluded with the following statement: "*We speculate that the symmetry effect of Coriolis coupling can appear in quantum mechanical analysis of the model.*"

Several accurate quantum calculations of the vibrational states in ozone molecule are available from literature,[29,30,31,32] but those are restricted to the ground rotational state ($J = 0$) and the first excited rotational state ($J = 1$ of positive parity) where the Coriolis coupling does not occur. Systematic studies of the rotationally excited states of ozone were conducted over the years by Tyuterev and coworkers using the method of effective Hamiltonian (see Ref. 33 and references therein). Their approach gives valuable interpretation of the experimental spectra, and, also permits to validate or even adjust the potential energy surface (PES) but, due to semi-empirical nature of their Hamiltonian, the method remains accurate only in a limited part of spectrum of given molecule, which restrains its predictive capability.

The first entirely general quantum calculation of the rotational-vibrational states in symmetric and asymmetric ozone molecules $^{16}O^{16}O^{18}O$ and $^{16}O^{18}O^{16}O$ with the Coriolis coupling terms included was published just recently.[11] As mentioned above, both the calculations themselves and the assignment of these states were challenging, so only the lowest 100 ro-vibrational states (for $^{16}O^{16}O^{18}O$ and $^{16}O^{18}O^{16}O$ combined) were computed, assigned and reported, up to only $J = 5$. This first step is encouraging, but now the question is how to push such calculations up to the level sufficient for the theoretical prediction of thermal rate coefficients for several isotopic variants of the ozone forming recombination reaction $O + O_2 \rightarrow O_3$.

Our experience with this reaction tells us that this task would be extremely demanding, since all states up to the dissociation threshold have to be computed (on the order of 300 vibrational states per rotational state of each isotopomer), plus scattering resonances above the dissociation threshold (say within a $3kT$ range), for the values of angular momentum up to at least $J = 50$. One can certainly argue that the role of the Coriolis coupling is very likely to be significant for these high levels of rotational and vibrational excitation. The practical question is how to conduct such calculations in the most efficient and accurate way.

One good idea is to switch from Jacobi coordinates used in Ref. 11 to the hyper-spherical coordinates[34,35,36,37,38] that are better suited for this system, for a number of reasons. First of all, hyper-spherical coordinates have a somewhat simpler form of the Hamiltonian operator and thus are more efficient numerically. Second, they fully exploit the symmetry of the ozone molecule and allow the treatment of both isotopomers of ozone, $^{16}O^{16}O^{18}O$ and $^{16}O^{18}O^{16}O$, on equal footing, covering all wells on the global PES with the same grids and/or basis sets. Although Jacoby coordinates also exploit full symmetry of the global PES, an incomplete basis set in Jacoby coordinates will result in different accuracies for the two isotopomers. Finally, hyper-spherical coordinates facilitate assignment of the computed ro-vibrational states.

However, within the hyper-spherical approach there are still several options worth exploring. Namely, it appears that different authors utilize different ways of positioning the *z*-axis relative to the molecular plane, which results in rather different structures of the corresponding Hamiltonian matrices.[34,35,36] One of our goals here is to offer a comprehensive catalogue of these possibilities, with clear analysis of the advantages and disadvantages of each choice. This is important since some of these options appear to be better suited for those molecules that are closer to the limit of a prolate symmetric top, while other options are more appropriate for the molecules that are closer to the limit of an oblate symmetric top. A concise summary of these possibilities has never been collected in one paper, to the best of our knowledge, which would certainly be useful for a meaningful application of this methodology to a broad range of molecules. These theoretical issues are addressed in Section II of the paper. Practical application of this methodology to the isotopically substituted ozone molecules $^{16}O^{16}O^{18}O$ and $^{16}O^{18}O^{16}O$ is presented in Section III, where the comparison of our data with those previously published[11] is also given. The conclusions and the prospects for determining the origin of mass-independent fractionation of oxygen isotopes, using this methodology, are outlined in Section IV.

## II. THEORY

Three usual Euler angles $(\alpha, \beta, \gamma)$ and three hyper-spherical coordinates $(\rho, \theta, \varphi)$ constitute the six degrees of freedom needed for the description of the coupled rotational-vibrational motion of a triatomic molecule. Qualitatively, the hyper-radius $\rho$ describes a "breathing" vibration mode, also known as symmetric stretch, whereas the hyper-angles $\theta$ and $\varphi$ correspond to bending and asymmetric-stretching modes of a triatomic.[39,40,41] In these coordinates the exact rotational-vibrational Hamiltonian operator is expressed in the following form:[13]

$$\hat{H} = \hat{H}_{\text{vib}} + \hat{T}_{\text{rot}} + \hat{T}_{\text{cor}} \tag{1}$$

The vibrational part of the Hamiltonian is separable in $\rho$, $\theta$ and $\varphi$ and includes, besides the potential energy surface $V_{\text{pes}}$, what we call the "extra-potential" term $V_{\text{ext}}$:

$$\hat{H}_{\text{vib}} = \hat{T}_\rho + \hat{T}_\theta + \hat{T}_\varphi + V_{\text{ext}} + V_{\text{pes}}(\rho, \theta, \varphi) \tag{2}$$

This simple form of the vibrational Hamiltonian enables efficient implementation of the sequential diagonalization-truncation method.[13] Actual expressions for these operators are:

$$\hat{T}_\rho = -\frac{\hbar^2}{2\mu}\frac{\partial^2}{\partial \rho^2} \tag{3}$$

$$\hat{T}_\theta = -\frac{2\hbar^2}{\mu \rho^2}\frac{\partial^2}{\partial \theta^2} \tag{4}$$

$$\hat{T}_\varphi = -\frac{2\hbar^2}{\mu \rho^2 \sin^2\theta}\frac{\partial^2}{\partial \phi^2} \tag{5}$$

$$V_{\text{ext}} = -\frac{\hbar^2}{2\mu\rho^2}\left(\frac{1}{4} + \frac{4}{\sin^2 2\theta}\right) \tag{6}$$

where

$$\mu = \sqrt{\frac{m_1 m_2 m_3}{m_1 + m_2 + m_3}} \tag{7}$$

is a three-atom reduced mass.

## A. Alternative choices of z-axis

Expressions for the rotational term $\hat{T}_{\text{rot}}$ and the rotation-vibration interaction (Coriolis) term $\hat{T}_{\text{cor}}$ depend on the choice of z-axis. In general, the adiabatically-adjusting principal-axes hyper-spherical (APH) coordinates employ three principal axes of inertia of the instantaneous geometry of the molecule (of its shape defined by $\rho$, $\theta$ and $\varphi$) to describe its rotational motion. Pack and Parker[34] have chosen, as their z-axis, one of the principal axes that lies in the molecular plane and follows the large-amplitude vibrational motion of the molecule as it distorts toward two-body dissociation (*e.g.*, O₃ → O + O₂). Then, for z in the plane:

$$\hat{T}_{\text{rot}} = A\hat{J}_x^2 + B\hat{J}_y^2 + C\hat{J}_z^2 \tag{8}$$

$$\hat{T}_{\text{cor}} = 4B \cos\theta \left(i\hbar \frac{\partial}{\partial \varphi}\right) \hat{J}_y \tag{9}$$

Note that according to their convention the *y*-axis is perpendicular to the molecular plane, so the Coriolis term couples $\hat{J}_y$ with $\partial/\partial\varphi$, which describes angular momentum of the pseudo-rotational motion (the so-called "internal rotation" of the molecule due to its vibration along $\varphi$). Expressions of the effective rotational constants of such a fluid rotor, as opposed to a rigid rotor, are listed in Table 1. Note that they depend on both $\rho$ and $\theta$, but not on $\varphi$.

Table 1. Correspondence between components of the total angular momentum *J* and rotational constants *A*, *B* and *C* of the fluid rotor in APH coordinates.

| *z* in plane (Pack & Parker) | $z \perp$ plane (Johnson; Kendrick) | Corresponding rotational constant |
|---|---|---|
| $J_x$ | $J_y$ | $A^{-1} = \mu\rho^2(1 + \sin\theta)$ |
| $J_y$ | $J_z$ | $B^{-1} = 2\mu\rho^2 \sin^2\theta$ |
| $J_z$ | $J_x$ | $C^{-1} = \mu\rho^2(1 - \sin\theta)$ |

Alternatively, Johnson[36] and independently Kendrick[35] have chosen their *z*-axis as the principal axis perpendicular to the molecular plane. Then, for $z \perp$ to the plane:

$$\hat{T}_{\text{rot}} = A\hat{J}_y^2 + B\hat{J}_z^2 + C\hat{J}_x^2 \tag{10}$$

$$\hat{T}_{\text{cor}} = 4B \cos\theta \left(i\hbar \frac{\partial}{\partial \varphi}\right) \hat{J}_z \tag{11}$$

In this case the Coriolis term couples $\hat{J}_z$ with $\partial/\partial\varphi$. To avoid confusion, we do not rename the effective rotational constants $A$, $B$ and $C$, but in this case they correspond to $\hat{J}_y$, $\hat{J}_z$ and $\hat{J}_x$, respectively, as emphasized by Table 1.

## B. The limits of prolate and oblate symmetric tops

The limiting case of a symmetric-top rotor is methodologically important, even for an asymmetric-top rotor treated exactly without any approximations, since it is always useful to split the terms in Eqs. (8) and (10) such that

$$\hat{T}_{\text{rot}} = \hat{T}_{\text{sym}} + \hat{T}_{\text{asym}} \tag{12}$$

It should be stressed that this expression is exact, not an approximation (not yet), but is merely a convenient re-arrangement of terms within $\hat{T}_{rot}$.

Let's assume that three moments of inertia of the molecule are such that it is close to the limit of a *prolate* symmetric top (such as a weakly bound van der Waals complex O···O$_2$).[14,26,42] First consider the case of Pack and Parker[34] when the $z$-axis lies in the molecular plane as shown by a pictogram at the bottom of Figure 1. In this case $I_x \approx I_y > I_z$, which corresponds to $A \approx B < C$. One can easily check that, using the definition $\hat{J}^2 = \hat{J}_x^2 + \hat{J}_y^2 + \hat{J}_z^2$, the expression of Eq. (8) can be rewritten in the form of Eq. (12) with the following assignments (prolate, $z$ in the plane):

$$\hat{T}_{sym} = \frac{A+B}{2}\hat{J}^2 + \left(C - \frac{A+B}{2}\right)\hat{J}_z^2 \qquad (13)$$

$$\hat{T}_{asym} = \frac{A-B}{2}(\hat{J}_x^2 - \hat{J}_y^2) \qquad (14)$$

Alternatively, in the case of Johnson[36] and Kendrick[35] with the $z$-axis perpendicular to the molecular plane, as shown by the pictogram at the bottom of Figure 2, we should set $I_y \approx I_z > I_x$ (although we still have $A \approx B < C$, according to the definition of Table 1). Therefore, the expression of Eq. (10) can be rewritten in the form of Eq. (12) with the following definitions (prolate, $z \perp$ to plane):

$$\hat{T}_{sym} = \frac{A+B}{2}\hat{J}^2 + \left(C - \frac{A+B}{2}\right)\hat{J}_x^2 \qquad (15)$$

$$\hat{T}_{asym} = \frac{A-B}{2}(\hat{J}_y^2 - \hat{J}_z^2) \qquad (16)$$

These assignments are physically appealing, since we defined two useful characteristics of the molecule. The first characteristic is the average of two (approximately equal) rotational constants $A$ and $B$ of a nearly prolate top, sometimes called $\tilde{A} = (A+B)/2$, which appears in Eqs. (13) and (15). Since this $\tilde{A} < C$, the last term in Eqs. (13) and (15) is positive. The second characteristic is the difference of $A$ and $B$ in Eqs. (14) and (16), namely $(A-B)/2$ that serves as a measure of deviation from the limit of a prolate symmetric top. Indeed, in the limiting case of a perfectly symmetric prolate top, when $A = B$ exactly, we correctly obtain $\hat{T}_{asym} = 0$, and then simply $\hat{T}_{rot} = \hat{T}_{sym}$.

We also see that if the rotor is not perfectly symmetric but is extremely prolate, $A \approx B \ll C$, then the effect of the asymmetric term is expected to be small, $\hat{T}_{asym} \ll \hat{T}_{sym}$, and one can consider an approximation in which this term is neglected (the symmetric-top approximation). Importantly, in this case the Coriolis coupling term, proportional to the value of $B$ in Eq. (9) or Eq. (11), is also small and can be neglected as well: $\hat{T}_{\text{cor}} \ll \hat{T}_{\text{sym}}$.

Similar derivations can be conducted for a molecule in which the three moments of inertia are such that it is close to the limit of an *oblate* symmetric top (such as cyclic-$O_3$, high-energy isomer in the form of an equilateral triangle).[43,44] Again, in the case of Pack and Parker[34] shown by a pictogram at the bottom of Figure 3, with the $z$-axis chosen in the plane of the molecule, we have $I_x \approx I_z < I_y$ and thus $A \approx C > B$. Now, in order to convert the expression of Eq. (8) into the form of Eq. (12) we should set (oblate, $z$ in the plane):

$$\hat{T}_{\text{sym}} = \frac{C+A}{2}\hat{J}^2 + \left(B - \frac{C+A}{2}\right)\hat{J}_y^2 \tag{17}$$

$$\hat{T}_{\text{asym}} = \frac{C-A}{2}(\hat{J}_z^2 - \hat{J}_x^2) \tag{18}$$

Alternatively, in the case of Johnson[36] and Kendrick[35] with $z$-axis perpendicular to the molecular plane as shown by the pictogram at the bottom of Figure 4, we set $I_x \approx I_y < I_z$, which still corresponds to $A \approx C > B$, and rearrange the terms of Eq. (10) as follows (oblate, $z \perp$ to plane):

$$\hat{T}_{\text{sym}} = \frac{C+A}{2}\hat{J}^2 + \left(B - \frac{C+A}{2}\right)\hat{J}_z^2 \tag{19}$$

$$\hat{T}_{\text{asym}} = \frac{C-A}{2}(\hat{J}_x^2 - \hat{J}_y^2) \tag{20}$$

We see that for a nearly oblate top the average of the two similar rotational constants is $\tilde{A} = (A+C)/2$. The $\tilde{A}$ appears in Eqs. (17) and (19) and we see that the last term in each of these formulae is negative, since this $\tilde{A} > B$. Deviation from the limit of oblate symmetric top is measured by $(C-A)/2$ in Eqs. (18) and (20). The limiting case of a perfectly symmetric oblate top, $A = C$, gives $\hat{T}_{\text{asym}} = 0$ and leads to $\hat{T}_{\text{rot}} = \hat{T}_{\text{sym}}$, as expected. If the rotor is extremely oblate, such as $A \approx C \gg B$, then we can again claim that the Coriolis coupling term, proportional to the value of $B$ in Eq. (9) or Eq. (11), is relatively small and can probably be neglected: $\hat{T}_{\text{cor}} \ll \hat{T}_{\text{sym}}$. However, in this case we can't anymore assume that the effect of the asymmetric term is also small, since both $A$ and $C$ are large and their difference is not necessarily small relative to $\tilde{A}$ and/or $B$. From these simple considerations it becomes clear that one should be careful when applying the symmetric-top approximation to the case of a nearly symmetric oblate top. This issue is further explored in the next subsection by computing matrix elements of these operators.

## C. Rotational-Vibrational wavefunctions

Each of the eigenstates (wave functions) $F_M^{Jpk}(\rho, \theta, \varphi, \alpha, \beta, \gamma)$ of the coupled rotational-vibrational Hamiltonian can be expressed in the following form:

$$F_M^{Jpk} = \sum_{K=0,1}^{J} \Psi_K^{Jpk}(\rho, \theta, \varphi) \widetilde{D}_{KM}^{Jp}(\alpha, \beta, \gamma) \tag{21}$$

where for any given $J$ the sum is over the modified Wigner functions $\widetilde{D}_{KM}^{Jp}$ of given inversion parity $p$, labeled by $K$. These functions are defined as normalized combinations of regular Wigner functions $D_{KM}^{J}$:

$$\widetilde{D}_{KM}^{Jp} = \sqrt{\frac{2J+1}{16\pi^2(1+\delta_{K0})}} \left[ D_{KM}^{J}(\alpha, \beta, \gamma) + (-1)^{J+K+p} D_{-KM}^{J}(\alpha, \beta, \gamma) \right] \tag{22}$$

The values of $p = 0$ and $p = 1$ generate two possible superpositions, one "in phase" and one "out of phase", except that in the case of $K = 0$ only the in-phase superposition is possible. For even $J$ the term with $K = 0$ contributes only to $p = 0$, while for odd $J$ the term with $K = 0$ contributes only to $p = 1$. In these cases, there are $J + 1$ terms in the sum of Eq. (21), with $K$ varied through the range $0 \le K \le J$. In the remaining cases, there are only $J$ terms in the sum of Eq. (21), with $K$ values in the range $1 \le K \le J$.

The index $k$ in $\widetilde{\Psi}_M^{Jpk}$ of Eq. (21) labels ro-vibrational eigenstates, within given $J$ and $p$. These are defined by a set of vibrational wave functions $\Psi_K^{Jpk}(\rho, \theta, \varphi)$, $0 \le K \le J$ for each $i$. These are determined numerically using an efficient sequential diagonalization-truncation approach that combines a symmetry-adapted FBR in $\varphi$, with a constant step DVR in $\theta$, and an optimized grid DVR along $\rho$.[13,45] Such an approach is numerically efficient and the readers are encouraged to become familiar with this earlier work.[13]

Note that the Wigner functions $\widetilde{D}_{KM}^{Jp}$ are eigenfunctions of both $\hat{J}^2$ and $\hat{J}_z^2$ with the following eigenvalues:[46]

$$\langle \widetilde{D}_{KM}^{Jp} | \hat{J}^2 | \widetilde{D}_{K'M}^{Jp} \rangle = \hbar^2 J(J+1) \tilde{\delta}_{KK'} \tag{23}$$

$$\langle \widetilde{D}_{KM}^{Jp} | \hat{J}_z^2 | \widetilde{D}_{K'M}^{Jp} \rangle = \hbar^2 K^2 \widetilde{\delta}_{KK'} \tag{24}$$

where $\widetilde{\delta}_{KK'}$ is similar to the Kronecker symbol $\delta_{KK'}$, except that $\widetilde{\delta}_{KK'}$ can be zero even in the case $K = K' = 0$ if $J + p$ is odd, namely:

$$\widetilde{\delta}_{KK'} = \begin{cases} \delta_{KK'} & \text{if } KK' \neq 0, \\ \delta_{(-1)^{J+p},1} & \text{if } KK' = 0. \end{cases} \tag{25}$$

This property will simplify the matrix elements of $\hat{T}_{\text{sym}}$ in Eqs. (13), (15), (17) and (19). However, the functions $\widetilde{D}_{KM}^{Jp}$ are not eigenfunctions of $\hat{J}_x^2$ and $\hat{J}_y^2$ in Eqs. (15) and (17) for $\hat{T}_{\text{sym}}$, neither are they eigenfunctions of $\hat{J}_x^2 - \hat{J}_y^2$, $\hat{J}_y^2 - \hat{J}_z^2$, and $\hat{J}_z^2 - \hat{J}_x^2$ in Eqs. (14), (16), (18) and (20) for $\hat{T}_{\text{asym}}$. These matrix elements are derived in Appendix A and are used in the next sub-section to derive the matrix elements of $\hat{T}_{\text{sym}}$, $\hat{T}_{\text{asym}}$ and $\hat{T}_{\text{cor}}$ for all of the cases introduced above. Since we will focus on the block-structure of the Hamiltonian matrix, with the blocks labeled by $K$, we will make all other indexes ($J$, $M$, $p$ and $k$) implicit in the functions $\Psi_K^{Jpk}$ and $\widetilde{D}_{KM}^{Jp}$ and will omit them for clarity.

Note also that in Eqs. (21)-(25) we used a generic symbol $K$ for the projection of $J$ onto the molecule-fixed axis $z$. In what follows we will use the symbol $\Lambda$ for the projection of $J$ onto the $z$-axis chosen to lie in the plane of the molecule, following the notation of Pack and Parker.[34] But we will use symbol $\Omega$ for the projection of $J$ onto the $z$-axis chosen to be perpendicular to the plane of the molecule, following the notation of Kendrick.[36]

### D. Matrix elements for prolate top

For the case of the $z$-axis *in the plane* of the molecule, for $\hat{T}_{\text{sym}}$ of Eq. (13) we obtain:

$$\langle \Psi_\Lambda \widetilde{D}_\Lambda | \hat{T}_{\text{sym}} | \Psi_{\Lambda'} \widetilde{D}_{\Lambda'} \rangle$$

$$= \langle \Psi_\Lambda | \frac{A+B}{2} | \Psi_{\Lambda'} \rangle \langle \widetilde{D}_\Lambda | \hat{J}^2 | \widetilde{D}_{\Lambda'} \rangle + \langle \Psi_\Lambda | C - \frac{A+B}{2} | \Psi_{\Lambda'} \rangle \langle \widetilde{D}_\Lambda | \hat{J}_z^2 | \widetilde{D}_{\Lambda'} \rangle$$

$$= \hbar^2 J(J+1) \langle \Psi_\Lambda | \frac{A+B}{2} | \Psi_{\Lambda'} \rangle \widetilde{\delta}_{\Lambda\Lambda'} + \hbar^2 \Lambda^2 \langle \Psi_\Lambda | C - \frac{A+B}{2} | \Psi_{\Lambda'} \rangle \widetilde{\delta}_{\Lambda\Lambda'} \tag{26}$$

From this formula one can see that in this case the symmetric-top rotor term contributes only to the diagonal blocks of the matrix. In the schematic of Figure 1 these blocks are labelled by letter "S". Note, that here and further in the text the matrix elements are computed not with the function $\Psi_\Lambda$ itself, but rather with the basis functions of its expansion. The details of calculation of these vibrational integrals will be discussed elsewhere.

For $\hat{T}_{asym}$ of Eq. (14) we obtain:

$$\langle\Psi_\Lambda\widetilde{D}_\Lambda|\hat{T}_{asym}|\Psi_{\Lambda'}\widetilde{D}_{\Lambda'}\rangle = \langle\Psi_\Lambda|\frac{A-B}{2}|\Psi_{\Lambda'}\rangle\langle\widetilde{D}_\Lambda|\hat{J}_x^2 - \hat{J}_y^2|\widetilde{D}_{\Lambda'}\rangle = \frac{\hbar^2}{4}U_{\Lambda\Lambda'}\langle\Psi_\Lambda|A-B|\Psi_{\Lambda'}\rangle \quad (27)$$

where

$$U_{\Lambda\Lambda'} = \frac{1}{\sqrt{(1+\delta_{\Lambda 0})(1+\delta_{\Lambda' 0})}}(\lambda_+(J,\Lambda)\lambda_+(J,\Lambda+1)\delta_{\Lambda,\Lambda'-2}$$
$$+ \lambda_+(J,\Lambda')\lambda_+(J,\Lambda'+1)\delta_{\Lambda,\Lambda'+2} + (-1)^{J+\Lambda+p}\lambda_+(J,\Lambda'-1)\lambda_+(J,\Lambda'-2)\delta_{\Lambda,2-\Lambda'}) \quad (28)$$

$$\lambda_\pm(J,\Lambda) = \sqrt{(J\pm\Lambda+1)(J\mp\Lambda)} \quad (29)$$

Details of the derivation of the matrix $U_{\Lambda\Lambda'}$ and additional discussion of its structure can be found in Appendix A. As one can see from Eq. (28), $U_{\Lambda\Lambda'}$ couples the blocks of the Hamiltonian matrix with $\Delta\Lambda = \pm 2$, but also contributes to one diagonal block $\Lambda = \Lambda' = 1$. Schematically these blocks are indicated by letter "A" in Figure 1.

For $\hat{T}_{cor}$ of Eq. (9) we obtain:

$$\langle\Psi_\Lambda\widetilde{D}_\Lambda|\hat{T}_{cor}|\Psi_{\Lambda'}\widetilde{D}_{\Lambda'}\rangle = \langle\Psi_\Lambda|4B\cos\theta\frac{d}{d\varphi}|\Psi_{\Lambda'}\rangle\langle\widetilde{D}_\Lambda|i\hbar\hat{J}_y|\widetilde{D}_{\Lambda'}\rangle$$
$$= 2\hbar^2 W_{\Lambda\Lambda'}\langle\Psi_\Lambda|B\cos\theta|d\Psi_{\Lambda'}/d\varphi\rangle \quad (30)$$

where

$$W_{\Lambda\Lambda'} = \frac{1}{\sqrt{(1+\delta_{\Lambda 0})(1+\delta_{\Lambda' 0})}}(\lambda_+(J,\Lambda)\delta_{\Lambda,\Lambda'-1} - \lambda_+(J,\Lambda')\delta_{\Lambda,\Lambda'+1}$$
$$+ (-1)^{J+\Lambda+p}\lambda_+(J,\Lambda'-1)\delta_{\Lambda,1-\Lambda'}) \quad (31)$$

Details of the derivation of the matrix $W_{\Lambda\Lambda'}$ and additional discussion of its structure can be found in Appendix A. As one can see from Eq. (31), $W_{\Lambda\Lambda'}$ couples the blocks of the Hamiltonian matrix with $\Delta\Lambda = \pm 1$. It means that only the first upper and first lower off-diagonal blocks of the Hamiltonian matrix are affected, as indicated by letter "C" in Figure 1.

For the case of *z*-axis *perpendicular to the plane* of the molecule, for $\hat{T}_{sym}$ of Eq. (15) we obtain:

$$\langle \Psi_\Omega \widetilde{D}_\Omega | \hat{T}_{\text{sym}} | \Psi_{\Omega'} \widetilde{D}_{\Omega'} \rangle$$

$$= \langle \Psi_\Omega | \frac{A+B}{2} | \Psi_{\Omega'} \rangle \langle \widetilde{D}_\Omega | \hat{J}^2 | \widetilde{D}_{\Omega'} \rangle + \langle \Psi_\Omega | C - \frac{A+B}{2} | \Psi_{\Omega'} \rangle \langle \widetilde{D}_\Omega | \hat{J}_x^2 | \widetilde{D}_{\Omega'} \rangle$$

$$= \hbar^2 J(J+1) \langle \Psi_\Omega | \frac{A+B}{2} | \Psi_{\Omega'} \rangle \tilde{\delta}_{\Omega\Omega'} + \frac{\hbar^2}{4} (S_{\Omega\Omega'} + U_{\Omega\Omega'}) \langle \Psi_\Omega | C - \frac{A+B}{2} | \Psi_{\Omega'} \rangle \tag{32}$$

Here the matrix $S_{\Omega\Omega'}$, defined in Eq. (A21) in Appendix A, affects only the diagonal blocks of the Hamiltonian matrix, i.e.: $\Delta\Lambda = 0$. Since both of the $S_{\Omega\Omega'}$ and $U_{\Omega\Omega'}$ terms appear in Eq. (32), $\hat{T}_{\text{sym}}$ contributes to both the main diagonal and second off-diagonal blocks of the Hamiltonian matrix, as indicated by letter "S" in Figure 2.

For $\hat{T}_{\text{asym}}$ of Eq. (16) we obtain:

$$\langle \Psi_\Omega \widetilde{D}_\Omega | \hat{T}_{\text{asym}} | \Psi_{\Omega'} \widetilde{D}_{\Omega'} \rangle = \langle \Psi_\Omega | \frac{A-B}{2} | \Psi_{\Omega'} \rangle \langle \widetilde{D}_\Omega | \hat{J}_y^2 - \hat{J}_z^2 | \widetilde{D}_{\Omega'} \rangle$$

$$= \frac{\hbar^2}{2} \left( (S_{\Omega\Omega'} - U_{\Omega\Omega'})/4 - \Omega^2 \tilde{\delta}_{\Omega\Omega'} \right) \langle \Psi_\Omega | A - B | \Psi_{\Omega'} \rangle \tag{33}$$

As in the case of Eq. (32), this matrix element contributes to the main diagonal and the second off-diagonal blocks of the Hamiltonian matrix, as indicated by letter "A" in Figure 2.

For $\hat{T}_{\text{cor}}$ of Eq. (11) we obtain:

$$\langle \Psi_\Omega \widetilde{D}_\Omega | \hat{T}_{\text{cor}} | \Psi_{\Omega'} \widetilde{D}_{\Omega'} \rangle = \langle \Psi_\Omega | 4B \cos\theta \frac{d}{d\varphi} | \Psi_{\Omega'} \rangle \langle \widetilde{D}_\Omega | i\hbar \hat{J}_z | \widetilde{D}_{\Omega'} \rangle$$

$$= 4i\hbar^2 \Omega \langle \Psi_\Omega | 4B \cos\theta | d\Psi_{\Omega'}/d\varphi \rangle \tilde{\delta}_{\Omega\Omega'} \tag{34}$$

In this case the Coriolis coupling term contributes to the main diagonal only, as indicated by letter "C" in Figure 2.

### E. Matrix elements for oblate top

For the case of $z$-axis *in the plane* of the molecule, for $\hat{T}_{\text{sym}}$ of Eq. (17) we obtain:

$$\langle \Psi_\Lambda \widetilde{D}_\Lambda | \hat{T}_{\text{sym}} | \Psi_{\Lambda'} \widetilde{D}_{\Lambda'} \rangle = \langle \Psi_\Lambda | \frac{C+A}{2} | \Psi_{\Lambda'} \rangle \langle \widetilde{D}_\Lambda | \hat{J}^2 | \widetilde{D}_{\Lambda'} \rangle + \langle \Psi_\Lambda | B - \frac{C+A}{2} | \Psi_{\Lambda'} \rangle \langle \widetilde{D}_\Lambda | \hat{J}_y^2 | \widetilde{D}_{\Lambda'} \rangle$$

$$= \hbar^2 J(J+1) \langle \Psi_\Lambda | \frac{C+A}{2} | \Psi_{\Lambda'} \rangle \tilde{\delta}_{\Lambda\Lambda'} + \frac{\hbar^2}{4} (S_{\Lambda\Lambda'} - U_{\Lambda\Lambda'}) \langle \Psi_\Lambda | B - \frac{C+A}{2} | \Psi_{\Lambda'} \rangle \tag{35}$$

For $\hat{T}_{\text{asym}}$ of Eq. (18) we obtain:

$$\langle \Psi_\Lambda \widetilde{D}_\Lambda | \hat{T}_{\text{asym}} | \Psi_{\Lambda'} \widetilde{D}_{\Lambda'} \rangle = \langle \Psi_\Lambda | \frac{C-A}{2} | \Psi_{\Lambda'} \rangle \langle \widetilde{D}_\Lambda | \hat{J}_z^2 - \hat{J}_x^2 | \widetilde{D}_{\Lambda'} \rangle$$

$$= \frac{\hbar^2}{2} \left( \tilde{\delta}_{\Lambda\Lambda'} \Lambda^2 - (S_{\Lambda\Lambda'} + U_{\Lambda\Lambda'})/4 \right) \langle \Psi_\Lambda | C - A | \Psi_{\Lambda'} \rangle \quad (36)$$

The Coriolis operator is the same as in the case of a prolate top, $z$-axis *in the plane*, Eq. (30). The overall structure of the matrix in this case is presented in Figure 3. The meaning of the letters is the same as in the case of a prolate top.

The operators $\hat{T}_{\text{sym}}$ and $\hat{T}_{\text{asym}}$ for the case of $z$-axis *perpendicular to the plane* are identical to the case of $z$-axis *in the plane* of a prolate top, except the names of rotational constants:

$$\langle \Psi_\Omega \widetilde{D}_\Omega | \hat{T}_{\text{sym}} | \Psi_{\Omega'} \widetilde{D}_{\Omega'} \rangle = \langle \Psi_\Omega | \frac{C+A}{2} | \Psi_{\Omega'} \rangle \langle \widetilde{D}_\Omega | \hat{J}^2 | \widetilde{D}_{\Omega'} \rangle + \langle \Psi_\Omega | B - \frac{C+A}{2} | \Psi_{\Omega'} \rangle \langle \widetilde{D}_\Omega | \hat{J}_z^2 | \widetilde{D}_{\Omega'} \rangle$$

$$= \hbar^2 J(J+1) \langle \Psi_\Omega | \frac{C+A}{2} | \Psi_{\Omega'} \rangle \tilde{\delta}_{\Omega\Omega'} + \hbar^2 \Omega^2 \langle \Psi_\Omega | B - \frac{C+A}{2} | \Psi_{\Omega'} \rangle \tilde{\delta}_{\Omega\Omega'} \quad (37)$$

$$\langle \Psi_\Omega \widetilde{D}_\Omega | \hat{T}_{\text{asym}} | \Psi_{\Omega'} \widetilde{D}_{\Omega'} \rangle = \langle \Psi_\Omega | \frac{C-A}{2} | \Psi_{\Omega'} \rangle \langle \widetilde{D}_\Omega | \hat{J}_x^2 - \hat{J}_y^2 | \widetilde{D}_{\Omega'} \rangle = \frac{\hbar^2}{4} U_{\Omega\Omega'} \langle \Psi_\Omega | C - A | \Psi_{\Omega'} \rangle \quad (38)$$

The Coriolis operator is the same as in the case of a prolate top, $z$-axis *perpendicular to the plane*, Eq. (34). The overall structure of the matrix in this case is presented in Figure 4. The meaning of the letters "S", "A" and "C" is the same as in the case of a prolate top.

## III. RESULTS AND DISCUSSION

### A. Comparison of different approaches

In the case of a prolate top, the advantage of the *z*-axis *lying in the plane* (Figure 1) is that the largest term ($\hat{T}_{\text{sym}}$) contributes solely to the main block diagonal, while smaller terms ($\hat{T}_{\text{cor}}$ and $\hat{T}_{\text{asym}}$) contribute to the first and second block off-diagonals, respectively. Such matrix structure provides a straightforward way of switching between the symmetric-top rotor approximation and the exact calculations. Indeed, as the molecular shape approaches the limit of a prolate symmetric top, the values in the off-diagonal blocks vanish and the matrix is effectively split into the individual $\Lambda$-blocks that can be diagonalized separately. This is exactly what we implemented for calculations reported in Ref. 13: the vibrational spectrum was computed separately for each pair of $(J, \Lambda)$ with the asymmetric-top rotor terms and Coriolis couplings neglected. As it was emphasized by Parker and Pack,[34] decoupling of different Λ-blocks requires neglecting both asymmetric-top rotor terms and Coriolis couplings, and constitutes one single approximation. In this case, neglecting only the asymmetric top rotor terms and keeping the Coriolis couplings, is not particularly useful since it would not lead to full decoupling of the diagonal blocks.

The case of *z*-axis *perpendicular to the molecule plane* (Figure 2), shows that $\hat{T}_{sym}$ contributes to diagonal blocks and the second off-diagonal blocks, which introduces substantial off-diagonal contribution. Because of that, even in the case of a perfectly symmetric top, one cannot neglect the values of the off-diagonal blocks and split the overall matrix into smaller pieces. Thus, in this case there is no way to implement the angular momentum decoupling. However, one advantage of this approach is that the blocks with even and odd values of $\Omega$ are not coupled with each other, which allows one to split the overall matrix into two blocks: one with the even values of $\Omega$ and one with the odd values of $\Omega$. It might be possible to do a similar separation in the case of the *z*-axis *in the plane* too, if the basis functions have distinct symmetry (the details of this are further discussed in section B of *Supplementary Information*).

In the case of an oblate top, the situation is completely opposite: now the choice of *z* ⊥ *to the plane* results in a simple matrix structure (Figure 4), where decoupling of different values of $\Omega$ can be achieved easily by neglecting the asymmetric top rotor terms, showing up in the second off-diagonal blocks (the contribution of the asymmetric term to $\Lambda = \Lambda' = 1$ block can be neglected as well). Moreover, there is no need to neglect the Coriolis couplings, since they contribute to the diagonal blocks only. In case of the *z*-axis lying *in the plane* (Figure 3), the contribution of the symmetric term to the off-diagonal blocks makes it impossible to decouple different values of $\Lambda$.

Another aspect worth considering is the cost of evaluation of the matrix elements. If every term is calculated independently, then the overall cost is proportional to the total number of terms (number of letters in Figures 1-4). For example, in the case of prolate top, *z in plane*, one would have to evaluate $J$ symmetric terms, $J-1$ Coriolis terms and $J-1$ asymmetric terms, a total of $3J-2$ terms. For the cases depicted in Figures 2, 3 and 4 the numbers are $5J-4$, $5J-5$ and $3J-1$, respectively, which makes *z in plane* optimal for the case of prolate top and *z* ⊥ *to the plane* optimal for the case of oblate top.

However, in practice, depending on the method of computation of the overlaps of the vibrational wavefunctions, it might be possible to reuse the intermediate values calculated for one term, to quickly calculate another term for the same set of $(\Lambda, \Lambda')$ or $(\Omega, \Omega')$. In this case the total cost of matrix element evaluations is proportional to the number of blocks that have at least one term in them. The total number of such blocks is $3J - 3$ for the case of *z in the plane* and $2J - 2$ for the case of *z ⊥ to the plane* for both prolate and oblate tops. This makes the choice of *z ⊥ to the plane* better in terms of the computational cost of evaluation of the matrix elements.

Finally, it might also be possible to recalculate every block quickly after the first block is calculated (or first few blocks). This could happen, for example, if the vibrational basis for different blocks is the same. In this case the difference between the cost of matrix evaluations in different approaches is not expected to be substantial.

## B. Application to ozone

In its equilibrium geometry (minimum energy point on the PES) the ozone molecule is close to a prolate-top rotor with $A = 0.446$ cm$^{-1}$, $B = 0.391$ cm$^{-1}$ and $C = 3.297$ cm$^{-1}$ (for the $^{16}O^{18}O^{16}O$ isotopomer).[47] As ozone dissociates, the shape of its rotor becomes even more prolate. Thus, a prolate symmetric top is often considered to be a reasonable approximation for the ozone molecule. For the reasons discussed above, and consistent with our previous work,[13] we have chosen to place the *z*-axis *in the plane* of molecule, which corresponds to Eqs. (26)-(30) for the matrix elements and Figure 1 for the block-structure of the matrix.

The potential energy surface used for the ro-vibrational calculations in this work was constructed by Dawes et al.[48] Technical details of our calculations are given in Section A of *Supplementary Information*. We carried out calculations for singly substituted isotopologue of ozone on the global PES which includes both symmetric $^{16}O^{18}O^{16}O$ and asymmetric $^{16}O^{16}O^{18}O$ isotopomers simultaneously, for rotational excitations up to $J = 5$, and for the rotational states of both values of parity $p$.

To begin with, we carried out calculations of the vibrational-rotational states of ozone in the symmetric-top rotor approximation with only diagonal blocks included ("S" in Figure 1), where both the asymmetric-top rotor terms and the Coriolis couplings were neglected. The results were identical to those found in our earlier work. Then, in one set of the intermediate calculations, in order to determine the role of asymmetry of the rotor, we added just the asymmetric-top rotor blocks to the matrix (only the "A" and "S" terms of Figure 1 were included in the Hamiltonian matrix) and we recomputed the vibrational-rotational states. Next, in the second set of the intermediate calculations, in order to determine the magnitude of the Coriolis effect alone, we added just the Coriolis coupling blocks to the matrix (only the "C" and "S" terms of Figure 1 were included in the Hamiltonian matrix) and we recomputed the vibrational-rotational states again. In the final set of exact calculations, we included all three types of blocks in the Hamiltonian matrix (the "S", "C" and "A" terms in Figure 1).

In Figure 5 we present the shifts of the energies of the ground vibrational state $(v_1, v_2, v_3) = (0,0,0)$ in $^{16}O^{18}O^{16}O$ due to inclusion of the asymmetric-top rotor term for the rotational excitation with $J = 5$. Here we see, first of all, a moderate negative shift by ~ 0.5 cm$^{-1}$ for the $\Lambda = 0$ state (parity is $p = 1$) and then two relatively large shifts of the $\Lambda = 1$ states, but in the opposite directions for two values of parity: positive shift for $p = 1$, and negative shift for $p = 0$. This creates a splitting of ~ 3.5 cm$^{-1}$. For $\Lambda = 2$ this splitting is reduced to ~ 0.5 cm$^{-1}$, in which case it is almost exclusively due to the positive shift of the $p = 1$ state, since the $p = 0$ state exhibits only a tiny shift. For $\Lambda = 3$ the shifts of the $p = 0$ and $p = 1$ states are both positive and small, which leads to a tiny splitting. For $\Lambda = 4$ and $\Lambda = 5$ the splittings of the ro-vibrational states of the two parities are vanishingly small.

In Figure 6 we present the shifts of energies of the ground vibrational state $(v_1, v_2, v_3) = (0,0,0)$ in $^{16}O^{18}O^{16}O$ due to inclusion of the Coriolis coupling term for the rotational excitation with $J = 5$. We see, first of all, that the Coriolis effect is an order of magnitude larger than the asymmetric-top rotor effect. For example, the shift of the $\Lambda = 0$ state is ~ 5 cm$^{-1}$. However, since the shifts are negative for *both* $p = 0$ and $p = 1$ parity states, the resultant splittings are of the same order of magnitude as before: close to 4 cm$^{-1}$ for $\Lambda = 1$, about 0.5 cm$^{-1}$ for $\Lambda = 2$, a tiny splitting for $\Lambda = 3$, and vanishingly small splittings for $\Lambda = 4$ and $\Lambda = 5$. Still, the shifts due to the Coriolis term are not small even for $\Lambda = 5$, which is close to negative 1 cm$^{-1}$ for both $p = 0$ and $p = 1$ parity states.

In order to understand the features of Figures 5 and 6, it is useful to analyze Eqs. (27) and (30), which define the asymmetric and Coriolis terms respectively. The magnitude of deviation from the energy of the symmetric top rotor is determined by the values of matrix elements of $\hat{T}_{asym}$ and $\hat{T}_{cor}$. One can see that, for the asymmetric term, the matrix elements are proportional to $U_{\Lambda\Lambda'}$ and $\frac{A-B}{4}$, while for the Coriolis term they are proportional to $W_{\Lambda\Lambda'}$ and $2B \cos\theta$. The structure of the matrixes $U_{\Lambda\Lambda'}$ and $W_{\Lambda\Lambda'}$, defined by Eqs. (28) and (31), is shown in Figures 7 and 8, respectively. The cases from $J = 0$ to $J = 3$ of both parities are depicted. In the cases when the $\Lambda = 0$ state is forbidden by symmetry the corresponding matrix elements are hatched. These pictures are analyzed in detail below.

For the equilibrium geometry of ozone, $\frac{A-B}{4} = 0.0138$ cm$^{-1}$ and $2B \cos \theta = 0.489$ cm$^{-1}$. Thus, according to Eqs. (27) and (30), the Coriolis coupling term is expected to be more important than the asymmetric top rotor term, at least for the low energy states (small $J$), which is indeed the case, as one can see from Figures 5 and 6. However, the values of $W_{\Lambda\Lambda'}$ grow only as $O(J)$, whereas the values of $U_{\Lambda\Lambda'}$ grow as $O(J^2)$, making the asymmetric-top rotor term more important for the highly excited rotational states (large $J$). It can also become more important for the excited vibrational states due to larger deviations from the equilibrium geometry. One can also see that the matrix elements $U_{\Lambda\Lambda'}$ and $W_{\Lambda\Lambda'}$ have their maximum values at $\Lambda = 0$ and decrease as $\Lambda$ increases, approaching the limit of $O(J)$ in the case of the asymmetric-top term and $O(\sqrt{J})$ in the case of the Coriolis term, but they never vanish. Because of that, the deviation from the symmetric top rotor limit would be the largest for small values of $\Lambda$, decrease as $\Lambda$ increases, but never reach zero, even when $\Lambda = J$. This is, indeed, what we see in Figures 5 and 6.

In contrast to the energy shifts, the splittings between the states of the two parities do not depend on the magnitudes of the matrix elements of $\hat{T}_{asym}$ and $\hat{T}_{cor}$ directly, but rather on the difference of their magnitudes for the cases of different parities. If one looks through the Eqs. (26)-(31), one will find that the parity affects two things only: first, whether or not the blocks with $\Lambda = 0$ are zero (and are excluded from the Hamiltonian, hatched area in Figures 7 and 8), and second, what is the sign of the diagonal block $\Lambda = \Lambda' = 1$ of the matrix $U_{\Lambda\Lambda'}$ (see Figure 7). This makes the $\Lambda = 1$ case the most susceptible to the splitting (at least for low values of $J$): in one parity it is coupled with the $\Lambda = 0$ state, in another parity it is not; in one parity the sign of the diagonal block $\Lambda = \Lambda' = 1$ of $U_{\Lambda\Lambda'}$ is positive, in another parity it is negative (with the same magnitude). The states with other values of $\Lambda$ do not experience such drastic differences due to the parity, thus, their splittings, being only an echo of the $\Lambda = 1$ splitting, decrease exponentially as $\Lambda$ increases and eventually vanish. As it was stated earlier, at high values of $J$ the asymmetric top rotor term is expected to take precedence over the Coriolis term. Thus, it is likely that for the high values of $J$, the splittings for the $\Lambda = 2$ state may even exceed those for $\Lambda = 1$, but not significantly.

In Figure 9 we present the shifts of energies of the ground vibrational state $(v_1, v_2, v_3) = (0,0,0)$ in $^{16}O^{18}O^{16}O$ for the rotational excitation with $J = 5$, due to inclusion of *both* the Coriolis coupling term and the asymmetric-top rotor term. Most importantly, this figure indicates that the energy shifts due to these two factors often occur in the *opposite* directions and thus partially cancel each other out with few exceptions (e.g. $\Lambda = 0$, and $\Lambda = 1$ and 2 with $p = 0$ for $J = 5$ where the shifts occur in the same directions). The value of the splitting for $\Lambda = 1$ is about 1 cm$^{-1}$, and it is only on the order of ~ 0.1 cm$^{-1}$ for $\Lambda = 2$. For $\Lambda \geq 3$ the splittings are negligible. However, the effect of the Coriolis coupling survives, since energies of all states are still reduced (relative to the symmetric-top rotor approximation) by a non-negligible shift. It varies in the range between negative 5 cm$^{-1}$ and 1 cm$^{-1}$ as the value of $\Lambda$ is increased from $\Lambda = 0$ to $\Lambda = 5$.

In the recent paper by Poirier and co-workers these splittings were named *the K-doubling effect*.[18] Alternatively, since we use symbol $\Lambda$ for the projection of $J$ onto $z$-axis, we could use the term *Λ-doubling* (although in this case one should be careful to avoid confusion, since this term is also used for the splitting of molecular levels due to interaction of molecular rotation with the orbital angular momentum of molecular electrons).[19] Or, this effect could also be named as *parity doubling*, since these splittings are caused by differences of rotational wavefunctions of two parities.

## C. Implications for ozone isotopomers

When all terms of the Hamiltonian matrix are included, our results show an excellent agreement with the results of the recent work by Poirier and co-workers.[11] Figure 10 plots the absolute values of the deviations of the state energies computed here relative to those reported in Ref. 11. These data include both symmetric $^{16}O^{18}O^{16}O$ and asymmetric $^{16}O^{16}O^{18}O$ isotopomers of ozone, combine the results of calculations with $J = 0, 1, 2, 3, 4$ and 5 for about 80 rotational-vibrational states of each parity, per each value of $J$ (about 850 states total). In Figure 10 each combination of $(J, p)$ is shown by its own color. As one can see from the picture, the differences of computed energies are on the order of $10^{-3}$ cm$^{-1}$ for the majority of states and on the order of $10^{-2}$ cm$^{-1}$ in the worst case, which matches the target accuracy of Poirier and coworkers. We found that the values of these differences depend on the vibrational character of the states $(v_1, v_2, v_3)$, but are relatively insensitive to the rotational quantum numbers $(J, \Lambda, p)$.

It should be stressed that the two sets of very similar results presented in Figure 10 (this work *vs.* Poirier and coworkers) were obtained independently by two groups without any communication, using different coordinates (hyper-spherical *vs.* Jacoby), employing two different codes (SpectrumSDT *vs.* ScalIT) and using different computer systems. The excellent agreement at low vibrational energies gave us enough confidence in the theory and the new code we developed to tackle a much more demanding problem – a large range of vibrational excitations. Namely, for each set of the rotational quantum numbers $(J, \Lambda, p)$ considered here we computed 600 vibrational states, 21600 coupled ro-vibrational states total. These new spectra cover roughly 90% of the covalent well of the ozone PES and stop just before the energy where the PES of ozone opens up toward a shallow plateau of the weak van der Waals interaction, followed by the bond breaking and dissociation onto $O + O_2$. (Calculations of the vibrational states in the remaining 10% of the energy range are also possible, but this would require a significant expansion of the $\rho$-grid, which is beyond the scope of this paper focused mostly on the rotation-vibrational coupling. Large-amplitude states near the threshold will be reported elsewhere, together with calculations of scattering resonances above the dissociation threshold). Figure 11 summarizes the energy progression of these ro-vibrational states for both symmetric $^{16}O^{18}O^{16}O$ and asymmetric $^{16}O^{16}O^{18}O$ ozone up to $J = 5$. We see that these spectra extend up to about 1000 cm$^{-1}$ below the dissociation threshold for all values of $J$. These data, including state energies, vibrational symmetries (see Section E of *Supplementary Information*), parities $p$, isotopomer-specific assignments (symmetric $^{16}O^{18}O^{16}O$ *vs.* asymmetric $^{16}O^{16}O^{18}O$) and the weights of all $\Lambda$-components for each ro-vibrational coupled state, are available from the archive file included in the *Supplementary Information*.

Overall, the spectra we computed and assigned contain up to the 11 quanta of bending motion, 8 quanta of asymmetric stretch and 7 quanta of symmetric stretch. For comparison, in the work of Poirier and co-workers[11] for $J = 5$ the states with no more than 2 quanta of vibrational excitation in one mode were computed. We found that the assignments of the vibrational states in terms of the normal mode quantum numbers $(v_1, v_2, v_3)$ are relatively certain for the lower 100 vibrational states for each set of $(J, \Lambda, p)$ for both $^{16}O^{18}O^{16}O$ and $^{16}O^{16}O^{18}O$. For completeness, we provided these assignments in the Tables of Section G of the *Supplementary Information*.

Figure 12(a) summarizes the progressions of energies for the normal mode overtones, while Figure 12(b) represents the dependence of parity splittings (or $\Lambda$-doublings) on the number of quanta in these vibrational progressions. The data for both $^{16}O^{18}O^{16}O$ and $^{16}O^{16}O^{18}O$ are included, separately. From Figure 12 one can see that the value of splitting monotonically increases for the bending mode progression and monotonically decreases for the symmetric stretching mode progression of ozone. In contrast, for the asymmetric stretching mode progression of ozone the value of splitting first increases and then slowly decreases, remaining roughly the same through a broad range of vibrational excitations. This makes sense, since the asymmetric stretching motion, described by the hyper-angle $\varphi$, does not affect the rotational constants of the molecule (see Table 1) and thus is not expected to change significantly its rotational asymmetry, which in turn makes the value of splitting relatively insensitive to the excitation of the asymmetric stretch. In contrast, excitation of the bending mode (described by the angle $\theta$) increases the asymmetry of the rotor (see Table 1), and thus is expected to increase the value of splitting. This is exactly what we see in Figure 12. It can also be concluded that the symmetric stretching motion makes the rotor more symmetric, since the magnitude of the splitting is significantly reduced by the excitation of the symmetric stretching mode. Finally, from Figure 12 we can see that, overall, the values of splittings are larger in the symmetric $^{16}O^{18}O^{16}O$ than in the asymmetric $^{16}O^{16}O^{18}O$.

It should be noted that for Figure 12 we selected the states with the dominant contribution of $\Lambda = 1$, for which the splittings are the largest (see Figure 9 above). Similar dependencies for $\Lambda = 2$, where the magnitudes of splittings are much smaller, can be found in the Section F of the *Supplemental Information*. Qualitatively, the splittings of the $\Lambda = 2$ states follow the same trends as we can see for $\Lambda = 1$ in Figure 12. Thus, we can conclude that the values of splittings do not change dramatically through the range of vibrational excitations considered here.

## IV. CONCLUSIONS

In this work we analyzed in detail several alternative ways of including the asymmetric-top rotor term and the Coriolis couplings in the accurate variational calculations of coupled rotational vibrational states, using hyper-spherical coordinates for a triatomic molecule. Namely, one can choose to place the *z*-axis of the coordinate system either in the plane of the molecule, or perpendicular to it. In each case the theory can be formulated in the way appropriate for a prolate top rotor, or for an oblate top rotor. Thus, four specific cases were considered here, each characterized by a distinct structure of the Hamiltonian matrix. We found that each case has its own advantages and/or disadvantages, and we discussed those in detail. These can be more significant or less significant, depending on the choice of the vibrational basis set and on the way the matrix elements are integrated. However, the case of an oblate top with the *z*-axis in the plane of a molecule seems to have no advantages (within the scope of criteria considered here) and thus should be avoided. Two of the four cases seem to be more advantageous, since they lead to the simplest form of the Hamiltonian matrix: prolate top with the *z*-axis chosen in the plane of the molecule, and oblate top with the *z*-axis chosen perpendicular to the molecular plane. These two cases are also more flexible, since, beside the exact calculations with all rotational-vibrational coupling included, they also permit to implement the symmetric-top rotor approximation within the same formalism and computer code.

The version of theory for a prolate top with the *z*-axis in plane was applied to compute the rotational-vibrational states of singly-substituted ozone isotopomers, symmetric $^{16}O^{18}O^{16}O$ and asymmetric $^{16}O^{16}O^{18}O$, for the rotational excitations from $J = 0$ (non-rotation ozone) to $J = 5$. The range of vibrational excitations extends up to 5 quanta of excitations in one mode. First, we carried out the simplest calculations within the symmetric-top rotor approximation, and then we added the asymmetric-top rotor terms and the Coriolis coupling terms, one at a time, and finally all together. This was done for the methodological reason, in order to illuminate the effect of each term on the spectrum of rotational-vibrational states, and most importantly on the *K*-doubling, which is the splitting of energies for the states of two parities. We showed that for the low values of rotational excitation in ozone, the Coriolis coupling effect is about an order of magnitude stronger than the asymmetric top rotor effect (in terms of shifts from the symmetric top rotor limit). The splittings due to the Coriolis and the asymmetric-top rotor effects, however, were on the same order of magnitude, but occurred in the opposite directions. Overall, in the exact calculations with both effects included, the influence of the two phenomena would partially cancel out, leading to relatively small residual splittings (*K*-doublings). Predicted energies of states are found to be in excellent agreement with recently published work of Poirier and coworkers.[11]

The methodology and computer code developed here can be used for calculations of accurate rotational vibrational states using the hyper-spherical coordinates for any triatomic molecule, in order to quantify its spectroscopy near the bottom of the well, or to access its chemical reactivity near the bond-breaking threshold and above it. In particular, it would be important to determine the role of rotational-vibrational couplings in the recombination reaction that forms ozone, focusing on the isotope effect. This is not an easy task, since it would require calculations for different isotopomers and isotopologues of ozone, such as $^{16}O^{18}O^{16}O$, $^{16}O^{16}O^{18}O$, $^{18}O^{16}O^{18}O$, $^{16}O^{18}O^{18}O$, $^{16}O^{17}O^{16}O$, $^{16}O^{16}O^{17}O$, $^{17}O^{16}O^{17}O$, $^{16}O^{17}O^{17}O$, in a broad range of rotational excitations (up to $J = 50$) and vibrational excitations up to the dissociation threshold (up to 10 quanta in one mode). This work is in progress and will be reported elsewhere. The code developed here (SpectrumSDT) will be made available to community in future releases.

## SUPPLEMENTARY INFORMATION

Section A provides further technical details of the calculations carried out in this work.

Section B describes possible ways of decoupling of the Hamiltonian matrix by symmetry or parity of $\Omega$ for each of the four cases considered in this work.

Section C provides definitions of the rotational operators $\hat{J}_x, \hat{J}_y$ and $\hat{J}_z$ and Wigner function.

Section D demonstrates the structure of matrix S (Eq. (A21), similar to Figures 7 and 8).

Section E gives a short summary of symmetry constrains of the wave functions in ozone.

Section F demonstrates dependency of vibrational energies and parity splittings for $J = 5$ and $\Lambda = 2$.

Section G lists all ro-vibrational levels of $^{16}O^{18}O^{16}O$ and $^{16}O^{16}O^{18}O$ isotopomers computed in this work.

## CONFLICT OF INTERESTS

There are no conflicts of interest to declare.

## ACKNOWLEDGMENTS

Richard Dawes is gratefully acknowledged for sharing the PES of ozone. This research was supported by the NSF AGS program Grant No. AGS-1920523. We used resources of the National Energy Research Scientific Computing Center, which is supported by the Office of Science of the U.S. Department of Energy under Contract No. DE-AC02-5CH11231. IG acknowledges the support of Schmitt Fellowship at Marquette. AT and BKK acknowledge that part of this work was done under the auspices of the US Department of Energy under Project No. 20180066DR of the Laboratory Directed Research and Development Program at Los Alamos National Laboratory. Los Alamos National Laboratory is operated by Triad National Security,

**FIGURES**

Figure 1. Rotational block structure of the Hamiltonian matrix for a prolate-top rotor molecule with $z$-axis in the molecular plane. Letters S, A and C indicate contributions from symmetric-top rotor, asymmetric-top rotor and Coriolis coupling terms, respectively. Other blocks of the matrix are zero. The blocks are labelled by $\Lambda$ and $\Lambda'$, the value of projection of total angular momentum $J$ onto $z$-axis.

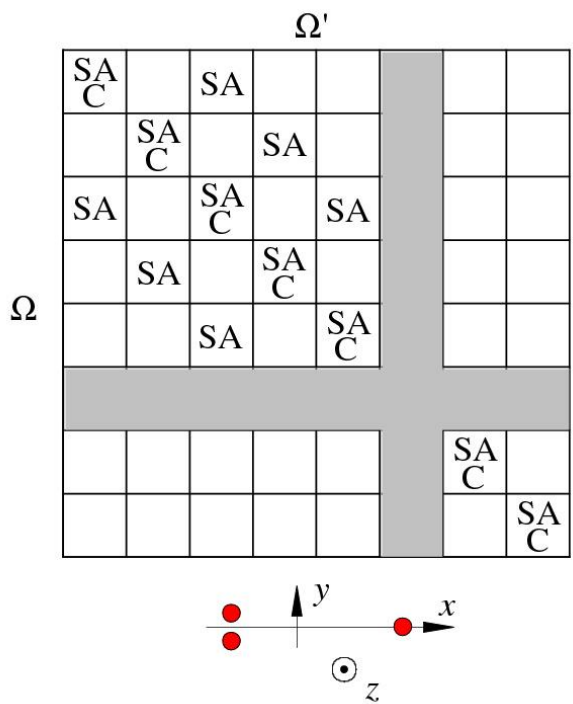

Figure 2. Same as in Figure 1, but for a prolate-top rotor molecule with $z$-axis perpendicular to the molecular plane. The blocks are labelled by $\Omega$ and $\Omega'$, the value of the projection of total angular momentum $J$ onto the $z$-axis.

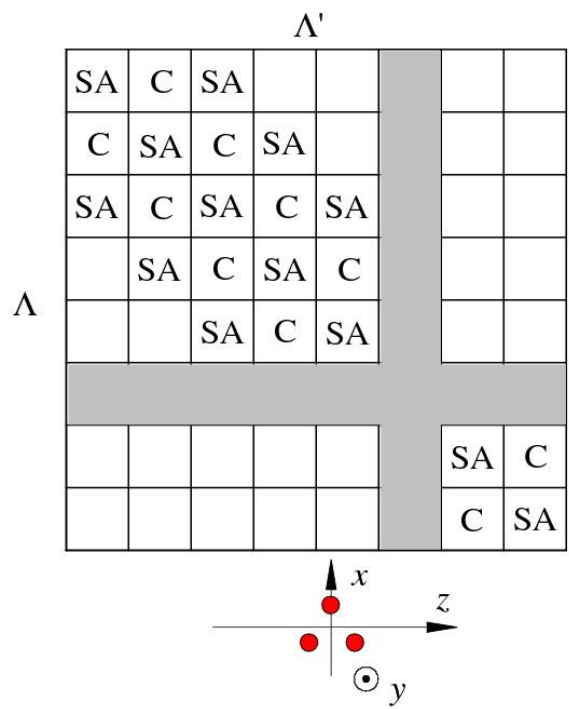

Figure 3. Same as in Figure 1, but for an oblate-top rotor molecule with *z*-axis in the molecular plane.

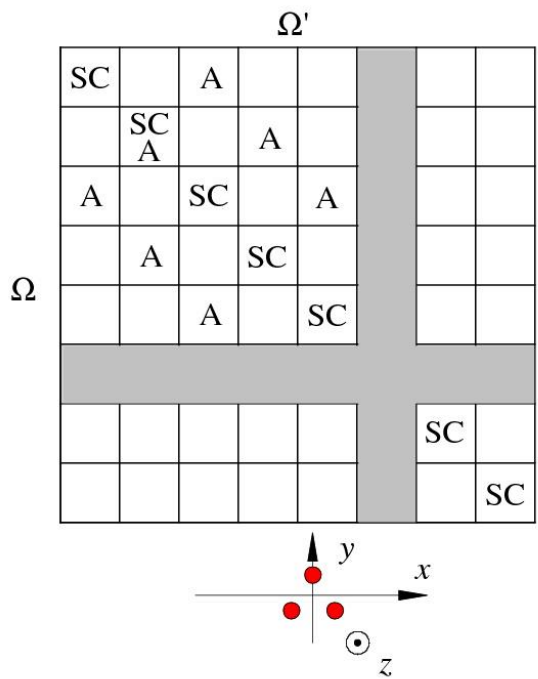

Figure 4. Same as in Figure 2, but for an oblate-top rotor molecule with *z*-axis perpendicular to the molecular plane.

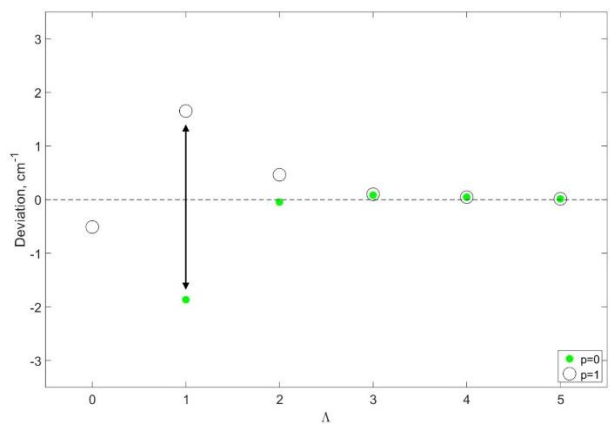

Figure 5. Deviations of the ground vibrational state of $^{16}O^{18}O^{16}O$ from the energies of a symmetric-top rotor due to the asymmetric-top rotor term for $J = 5$. The value of $\Lambda$ is plotted along the horizontal axis. The states of two different parities are denoted by color and symbol type. The magnitude of splitting (*K*-doubling) for $\Lambda = 1$ is indicated by a double arrow.

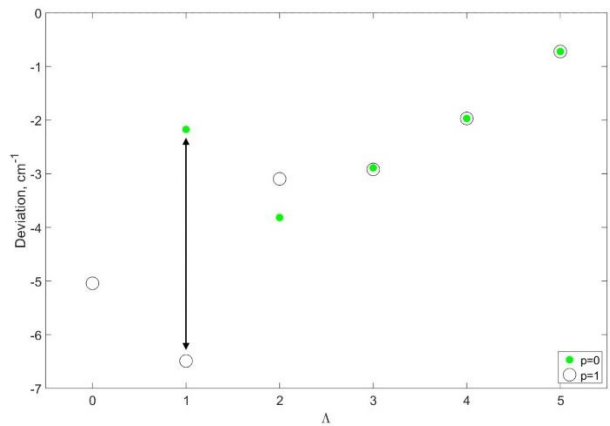

Figure 6. Same as in Figure 5, but for the deviations from energies of a symmetric-top rotor due to the Coriolis coupling term (alone, without the asymmetric-top rotor term).

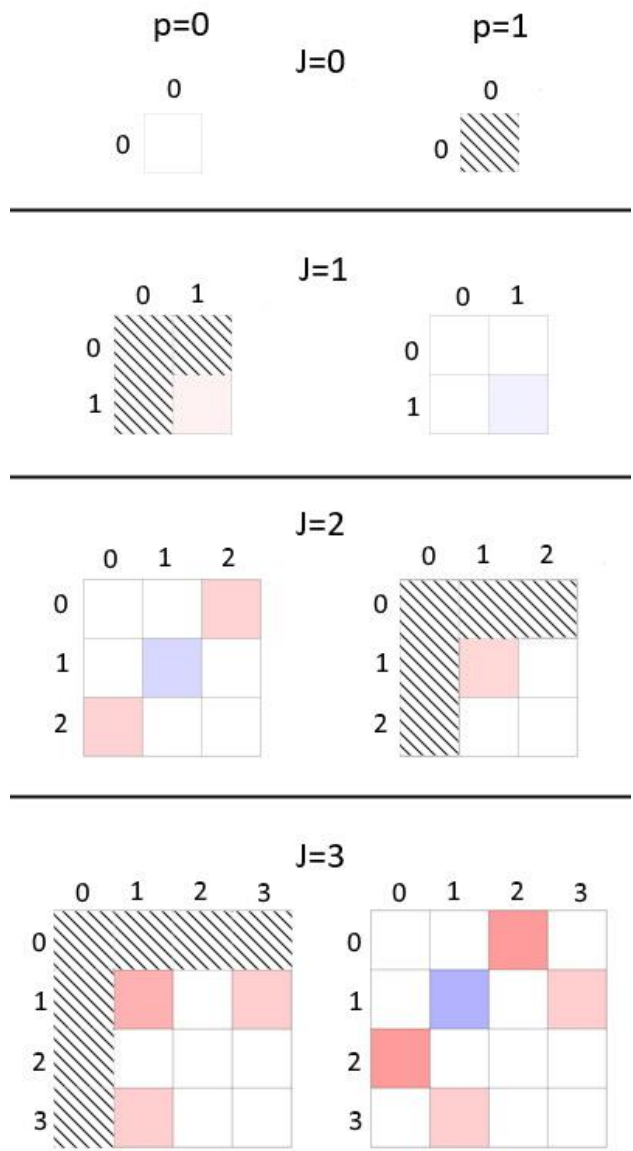

Figure 7. Block structure of the matrix $U_{\Lambda\Lambda'}$ for the rotational states from $J = 0$ to $J = 3$ (intuitive extrapolation to larger values of $J$ is relatively straightforward). Two parities are shown separately: $p = 0$ in the left column and $p = 1$ in the right column. As in Figure 1, the blocks are labelled by $\Lambda$ and $\Lambda'$. Color indicates magnitudes of matrix elements, with red being positive, blue being negative, and white being zero. When $J + p$ is odd, all states corresponding to $\Lambda = 0$ or $\Lambda' = 0$ are forbidden by symmetry and the corresponding blocks of the Hamiltonian matrix are excluded (hatched).

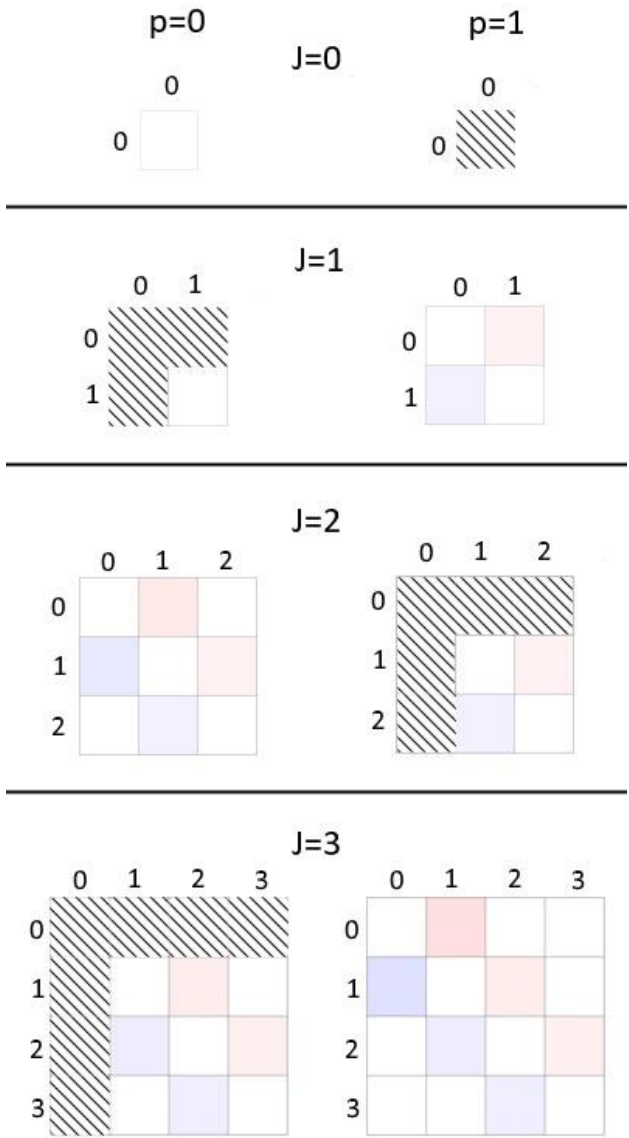

Figure 8. Same as in Figure 7, but for the matrix $W_{\Lambda\Lambda'}$.

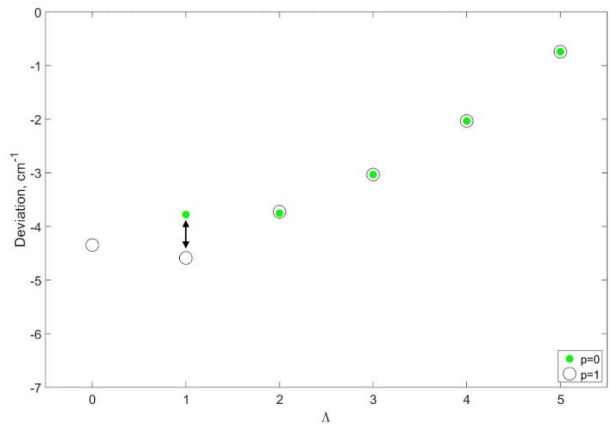

Figure 9. Same as in Figures 5 and 6, but for the case when both the asymmetric-top rotor term and the Coriolis coupling term are included in the Hamiltonian, which corresponds to the deviation of the exact rotational-vibrational state energies from the symmetric-top rotor approximation.

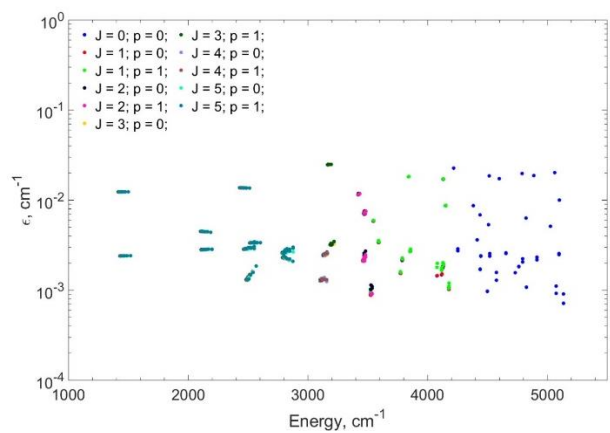

Figure 10. Absolute values of energy differences between the rotational-vibrational states computed here, and the corresponding states reported in reference 11 for $^{16}O^{18}O^{16}O$ and $^{16}O^{16}O^{18}O$. Individual colors are used for different parities $p$ and different values of angular momentum up to $J = 5$.

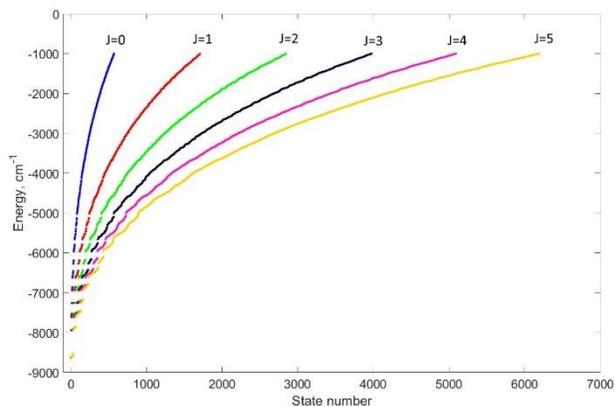

Figure 11. The progressions of energies of coupled ro-vibrational states up to $J = 5$ computed in this work for symmetric $^{16}O^{18}O^{16}O$ and asymmetric $^{16}O^{16}O^{18}O$ combined.

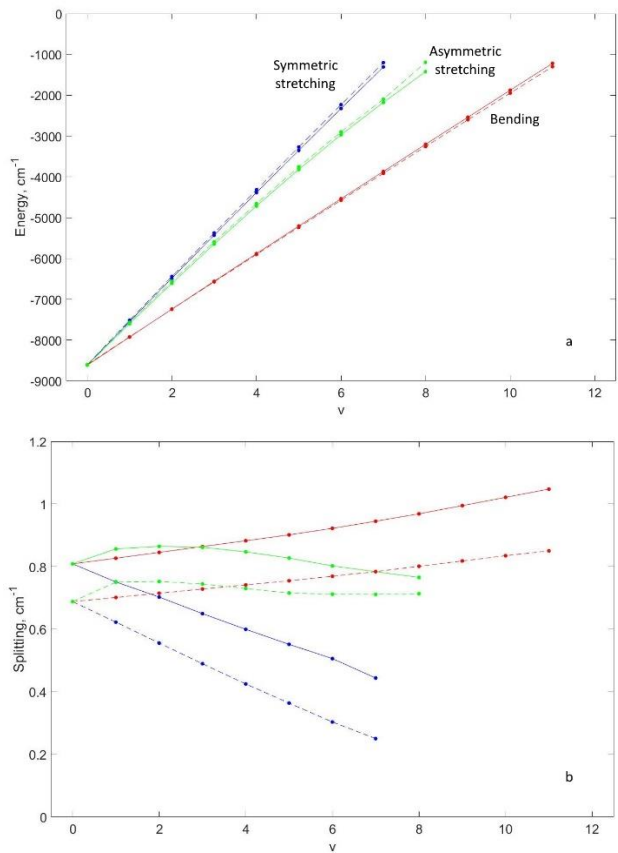

Figure 12. Evolution of energies and parity splittings for $J = 5$ and $\Lambda = 1$ as a function of number of vibrational quanta along the three normal modes of ozone. For each progression, the other two normal modes are not excited ($v = 0$). Solid and dashed lines correspond to symmetric $^{16}O^{18}O^{16}O$ and asymmetric $^{16}O^{16}O^{18}O$ ozone isotopomers, respectively. Progressions in the lower frame have the same colors as those in the upper frame.

## APPENDIX A. DERIVATION OF MATRIXES U, S AND W

The following integral can be rewritten in terms of raising and lowering operators:

$$\langle \widetilde{D}_K | \hat{J}_x^2 - \hat{J}_y^2 | \widetilde{D}_{K'} \rangle = \frac{1}{2}(\langle \widetilde{D}_K | \hat{J}_+^2 | \widetilde{D}_{K'} \rangle + \langle \widetilde{D}_K | \hat{J}_-^2 | \widetilde{D}_{K'} \rangle) \quad (A1)$$

where raising and lowering operators, $\hat{J}_+$ and $\hat{J}_-$, are defined as:[34]

$$\hat{J}_+ = \hat{J}_x - i\hat{J}_y \quad (A2)$$

$$\hat{J}_- = \hat{J}_x + i\hat{J}_y \quad (A3)$$

Or, the other way around:

$$\hat{J}_x = \frac{\hat{J}_+ + \hat{J}_-}{2} \quad (A4)$$

$$\hat{J}_y = \frac{\hat{J}_- - \hat{J}_+}{2i} \quad (A5)$$

Definition of $\hat{J}_x, \hat{J}_y$ and $\hat{J}_z$ in terms of elementary operators is given in section C of *Supplementary Information*. Application of $\hat{J}_\pm$ to a Wigner function is given by:[35]

$$\hat{J}_\pm D_K = \hbar \lambda_\pm(J,K) D_{K\pm 1} \quad (A6)$$

$$\lambda_\pm(J,K) = \sqrt{(J \pm K + 1)(J \mp K)} \quad (A7)$$

where the functions $\lambda_\pm(J,K)$ have the following useful properties (directly from their definitions):

$$\lambda_+(J,-K) = \lambda_+(J,K-1) \quad (A8)$$

$$\lambda_-(J,K) = \lambda_+(J,K-1) \quad (A9)$$

An extended version of these relationships can be found in Eq. (173) of Ref. 34. The first term of Eq. (A1) is:

$$\langle \widetilde{D}_K | \hat{J}_+^2 | \widetilde{D}_{K'} \rangle = \frac{2J+1}{16\pi^2 \sqrt{(1+\delta_{K0})(1+\delta_{K'0})}}$$

$$\times \langle D_K + (-1)^{J+K+p} D_{-K} | \hat{J}_+^2 | D_{K'} + (-1)^{J+K'+p} D_{-K'} \rangle$$

$$= \frac{2J+1}{16\pi^2 \sqrt{(1+\delta_{K0})(1+\delta_{K'0})}} \left( \hbar^2 \lambda_+(J,K') \lambda_+(J,K'+1) \langle D_K | D_{K'+2} \rangle \right.$$

$$+ (-1)^{J+K+p} \hbar^2 \lambda_+(J,K') \lambda_+(J,K'+1) \langle D_{-K} | D_{K'+2} \rangle$$

$$+ (-1)^{J+K'+p} \hbar^2 \lambda_+(J,-K') \lambda_+(J,-K'+1) \langle D_K | D_{-K'+2} \rangle$$

$$\left. + (-1)^{J+K+p}(-1)^{J+K'+p} \hbar^2 \lambda_+(J,-K') \lambda_+(J,-K'+1) \langle D_{-K} | D_{-K'+2} \rangle \right) \quad (A10)$$

Normalization of Wigner function is given by:

$$\langle D_K | D_{K'} \rangle = \frac{8\pi^2}{2J+1} \delta_{K,K'} \tag{A11}$$

Plugging Eq. (A11) into Eq. (A10) one obtains:

$$\langle \widetilde{D}_K | \hat{J}_+^2 | \widetilde{D}_{K'} \rangle = \frac{\hbar^2}{2\sqrt{(1+\delta_{K0})(1+\delta_{K'0})}} \big( \lambda_+(J,K')\lambda_+(J,K'+1)\delta_{K,K'+2}$$
$$+ (-1)^{J+K+p} \lambda_+(J,K')\lambda_+(J,K'+1)\delta_{-K,K'+2}$$
$$+ (-1)^{J+K'+p} \lambda_+(J,-K')\lambda_+(J,-K'+1)\delta_{K,-K'+2}$$
$$+ (-1)^{J+K+p}(-1)^{J+K'+p} \lambda_+(J,-K')\lambda_+(J,-K'+1)\delta_{-K,-K'+2} \big) \tag{A12}$$

The second term of the sum is 0, since $\delta_{-K,K'+2} = 0$ for $K, K' \geq 0$. In the fourth term, $(-1)^{J+K+p}(-1)^{J+K'+p} = 1$ for all $K, K'$ such that $\delta_{-K,-K'+2} \neq 0$. Plugging these results back into Eq. (A12) and using Eq. (A8) we obtain the final expression for the first term in (A1):

$$\langle \widetilde{D}_K | \hat{J}_+^2 | \widetilde{D}_{K'} \rangle = \frac{\hbar^2}{2\sqrt{(1+\delta_{K0})(1+\delta_{K'0})}} \big( \lambda_+(J,K')\lambda_+(J,K'+1)\delta_{K,K'+2}$$
$$+ (-1)^{J+K'+p} \lambda_+(J,K'-1)\lambda_+(J,K'-2)\delta_{K,2-K'}$$
$$+ \lambda_+(J,K'-1)\lambda_+(J,K'-2)\delta_{K,K'-2} \big) \tag{A13}$$

In a very similar way one can derive an expression for the second term:

$$\langle \widetilde{D}_K | \hat{J}_-^2 | \widetilde{D}_{K'} \rangle = \frac{\hbar^2}{2\sqrt{(1+\delta_{K0})(1+\delta_{K'0})}} \big( \lambda_-(J,K')\lambda_-(J,K'-1)\delta_{K,K'-2}$$
$$+ (-1)^{J+K+p} \lambda_-(J,K')\lambda_-(J,K'-1)\delta_{K,2-K'}$$
$$+ \lambda_-(J,K'+1)\lambda_-(J,K'+2)\delta_{K,K'+2} \big) \tag{A14}$$

Using Eq. (A9), Eq. (A14) can be rewritten as:

$$\langle \widetilde{D}_K | \hat{J}_-^2 | \widetilde{D}_{K'} \rangle = \frac{\hbar^2}{2\sqrt{(1+\delta_{K0})(1+\delta_{K'0})}} \big( \lambda_+(J,K'-1)\lambda_+(J,K'-2)\delta_{K,K'-2}$$
$$+ (-1)^{J+K+p} \lambda_+(J,K'-1)\lambda_+(J,K'-2)\delta_{K,2-K'}$$
$$+ \lambda_+(J,K')\lambda_+(J,K'+1)\delta_{K,K'+2} \big) \tag{A15}$$

Substitution of Eqs. (A13) and (A15) into Eq. (A1) gives:

$$\langle \widetilde{D}_K | \hat{J}_x^2 - \hat{J}_y^2 | \widetilde{D}_{K'} \rangle$$

$$= \frac{\hbar^2}{4\sqrt{(1+\delta_{K0})(1+\delta_{K'0})}} \big( 2\lambda_+(J,K')\lambda_+(J,K'+1)\delta_{K,K'+2}$$

$$+ \big((-1)^{J+K'+p} + (-1)^{J+K+p}\big)\lambda_+(J,K'-1)\lambda_+(J,K'-2)\delta_{K,2-K'}$$

$$+ 2\lambda_+(J,K'-1)\lambda_+(J,K'-2)\delta_{K,K'-2} \big) \tag{A16}$$

For any $K, K'$ such that $\delta_{K,2-K'} \neq 0$ $(-1)^{J+K'+p} + (-1)^{J+K+p} = 2(-1)^{J+K+p}$, so Eq. (A16) can be simplified to:

$$\langle \widetilde{D}_K | \hat{J}_x^2 - \hat{J}_y^2 | \widetilde{D}_{K'} \rangle = \frac{\hbar^2}{2} U_{KK'} \tag{A17}$$

where $U_{KK'}$ is defined as:

$$U_{KK'} = \frac{1}{\sqrt{(1+\delta_{K0})(1+\delta_{K'0})}} \big( \lambda_+(J,K)\lambda_+(J,K+1)\delta_{K,K'-2}$$

$$+ \lambda_+(J,K')\lambda_+(J,K'+1)\delta_{K,K'+2}$$

$$+ (-1)^{J+K+p}\lambda_+(J,K'-1)\lambda_+(J,K'-2)\delta_{K,2-K'} \big) \tag{A18}$$

The first two terms of Eq. (A18) make equal contributions to the second upper and lower off-diagonal blocks, respectively. The last term affects blocks (0, 2), (1, 1) and (2, 0). Because of the last term, the values of blocks (0, 2) and (2, 0) can either be doubled (if $J + p$ is even) or become 0 (if $J + p$ is odd). The magnitude of the values of matrix elements is on the order of $O(J^2)$. The structure of matrix $U_{KK'}$ is shown in Figure 7.

Now consider the integral with $\hat{J}_x^2$ operator alone. Using Eq. (A4), this integral can be expressed as:

$$\langle \widetilde{D}_K | \hat{J}_x^2 | \widetilde{D}_{K'} \rangle = \frac{1}{4} \big( \langle \widetilde{D}_K | \hat{J}_+\hat{J}_- + \hat{J}_-\hat{J}_+ | \widetilde{D}_{K'} \rangle + \langle \widetilde{D}_K | \hat{J}_+^2 + \hat{J}_-^2 | \widetilde{D}_{K'} \rangle \big) \tag{A19}$$

In a way very similar to the derivation of Eq. (A13), the first term of Eq. (A19) can be transformed to:

$$\langle \widetilde{D}_K | \hat{J}_+\hat{J}_- + \hat{J}_-\hat{J}_+ | \widetilde{D}_{K'} \rangle = \hbar^2 S_{KK'} \tag{A20}$$

where $S_{KK'}$ is defined as:

$$S_{KK'} = \big(\lambda_+^2(J,K-1) + \lambda_+^2(J,K)\big)\tilde{\delta}_{KK'} \tag{A21}$$

The structure of matrix $S_{KK'}$ is shown in Section E of *Supplementary Information*. The matrix elements have magnitudes on the order of $O(J^2)$.

The second term of Eq. (A19) has already been considered in Eq. (A1). Plugging these results back into Eq. (A19) one obtains:

$$\langle \widetilde{D}_K | \hat{J}_x^2 | \widetilde{D}_{K'} \rangle = \frac{\hbar^2}{4} (S_{KK'} + U_{KK'}) \tag{A22}$$

Similarly:

$$\langle \widetilde{D}_K | \hat{J}_y^2 | \widetilde{D}_{K'} \rangle = \frac{\hbar^2}{4} (S_{KK'} - U_{KK'}) \tag{A23}$$

Next, consider $\langle \widetilde{D}_K | i\hbar \hat{J}_y | \widetilde{D}_{K'} \rangle$. This integral can be expressed through $\hat{J}_+$ and $\hat{J}_-$ as:

$$\langle \widetilde{D}_K | i\hbar \hat{J}_y | \widetilde{D}_{K'} \rangle = \frac{\hbar}{2} \left( \langle \widetilde{D}_K | \hat{J}_- | \widetilde{D}_{K'} \rangle - \langle \widetilde{D}_K | \hat{J}_+ | \widetilde{D}_{K'} \rangle \right) \tag{A24}$$

Following the footsteps of the derivation preceding Eq. (A13), one can show that

$$\langle \widetilde{D}_K | \hat{J}_+ | \widetilde{D}_{K'} \rangle = \frac{\hbar}{2\sqrt{(1 + \delta_{K0})(1 + \delta_{K'0})}} \big(\lambda_+(J, K') \delta_{K, K'+1}$$
$$+ (-1)^{J+K'+p} \lambda_+(J, K'-1) \delta_{K, 1-K'} - \lambda_+(J, K'-1) \delta_{K, K'-1}\big) \tag{A25}$$

and

$$\langle \widetilde{D}_K | \hat{J}_- | \widetilde{D}_{K'} \rangle = \frac{\hbar}{2\sqrt{(1 + \delta_{K0})(1 + \delta_{K'0})}} \big(\lambda_+(J, K'-1) \delta_{K, K'-1}$$
$$+ (-1)^{J+K+p} \lambda_+(J, K'-1) \delta_{K, 1-K'} - \lambda_+(J, K') \delta_{K, K'+1}\big) \tag{A26}$$

Plugging Eqs. (A25) and (A26) back into Eq. (A24) one obtains:

$$\langle \widetilde{D}_K | i\hbar \hat{J}_y | \widetilde{D}_{K'} \rangle = \frac{\hbar^2}{2} W_{KK'} \tag{A27}$$

where $W_{KK'}$ is defined as:

$$W_{KK'} = \frac{1}{\sqrt{(1 + \delta_{K0})(1 + \delta_{K'0})}} \big(\lambda_+(J, K) \delta_{K, K'-1} - \lambda_+(J, K') \delta_{K, K'+1}$$
$$+ (-1)^{J+K+p} \lambda_+(J, K'-1) \delta_{K, 1-K'}\big) \tag{A28}$$

The structure of matrix $W_{KK'}$ is shown in Figure 8. Just like in the case of $U_{KK'}$, the last term of Eq. (A28) either doubles or nullifies the blocks (0, 1) and (1, 0) depending on parity of $J + p$. The magnitude of the values of matrix elements is on the order of $O(J)$.